# Design, Construction and Test Arrangement of a Fast-Cycling HTS Accelerator Magnet

Henryk Piekarz, Jamie Blowers, Steven Hays and Vladimir Shiltsev

*Abstract*—Design, fabrication and assembly of a novel fast-cycling accelerator magnet is presented. A short-sample magnet is powered with a single-turn HTS cable capable to carry 80 kA current at 20 K and generate 1.75 T field in a 40 mm magnet gap. The applied conventional leads and the power supply, however, allow only for a sin-wave 24 kA, 20 Hz current limiting test magnet to a B-field of 0.5 T and to a maximum cycling rate of 20 T/s. The critical aspects of the cable construction and the splicing connection to the power leads are described. Tentative power losses of the proposed HTS accelerator magnet in a possible application for proton and muon accelerators are presented.

*Index Terms*— **Fast-cycling accelerator magnet, HTS power cable, Conventional power leads, SC cable power losses.**

## I. Introduction

THE new developments [1,2] in the high-energy particle physics call for the construction of synchrotrons capable to accelerate high-intensity muon and electron beams in a fast-cycling mode of operation. Due to a large scale of these synchrotrons, however, the accelerator magnets should be characterized by very low dynamic power losses caused by hysteresis and eddy currents to make the operations practical and cost-effective.

A super-ferric magnet is constructed of three components: (1) magnetic core, (2) superconducting power cable and (3) power leads connecting superconducting cable with the current supply at room temperature. The most significant dynamic power losses are those of the magnetic core and the superconducting cable but the power leads require careful consideration as well. The test magnet is designed to generate a 1.75 T field in the 40 mm gap. The conventional leads and magnet cable will allow carry current up to 24 kA, generating field of 0.5 T in the magnet gap.

## II. Magnet system design

### A. Magnetic Core

The eddy and hysteresis power losses in the core depend on core cross-section area and thickness of the laminations. In the iron dominated magnet the core cross-section is primarily determined by the size of the beam gap and the cable cavities.

The high current density in the superconductor minimizes the cable size thus requiring smaller cable cavities than with the resistive cable and reducing the overall cross-section area of the magnet core.

For the proposed fast-cycling magnets we use 104 μm thick $Fe_3\%Si$ laminations. Combination of the small cross-section area of the core with the ultra-thin laminations allows strongly reduce both hysteresis and eddy currents losses in the core.

In the fast-cycling operation the power loss in the conductor is primarily due to the sweeping magnetic field crossing its space. As indicated in [3,4] the effect of the core induced magnetic field can be much reduced with a conductor of a narrow rectangular shape stretching as much as possible from the top to the bottom wall of the cable cavity. The field crossing the conductor space can be then further suppressed by the optimization of the cable cavity width and the cable position in that cavity. The effect of cable position on the magnetic field in the cable cavity is shown in Figs.1 A, B, C, D. The cable placement between positions B and C (~ 20 mm range) is the most optimal one for suppressing the core induced field in the conductor space.

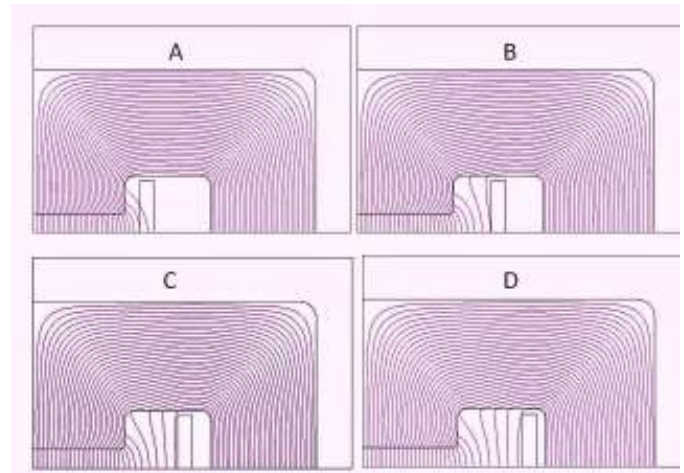

Fig. 1 Field in cable cavity at various cable horizontal positions. Magnetic field design is for 1.75 T in the beam gap. Active cable size is 8 mm wide x 60 mm high. The overall magnet core size is 32 cm (H) x 19 cm (V).

The test magnet core features beam gap of 100 mm (H) x 40 mm (V) and two cable gaps of 48 mm (H) x 65 mm (V) each. There are ~ 1150 104 μm laminations stacked in the 1.2 m long core. The cables are placed at about 5 mm off their cavity centers toward the outer core outer walls.

Manuscript received July 12, 2013. This work was supported in part by the Fermi Research Alliance, LLC under DOE Contract DE-AC02-07CH11359.

J. Blowers, S. Hays, H. Piekarz, and V. Shiltsev are with Fermi National Accelerator Laboratory, Batavia, IL 60510 USA (phone: 630-840-2105; fax: 630-840-6039; e-mail: hpiekarz@fnal.gov).

1PoCO-03                                                                                                                                          2*B. Superconducting Cable*

The HTS superconductor in the tape form is very suitable for fabrication a cable which is narrow but can be extended vertically as required for the core design in Fig. 1. For the test magnet the cable is constructed of six sub-cables stacked vertically, one on the top of other. Each sub-cable consists of twenty 3 m long 344C-2G tapes of 0.2 mm x 4.5 cross-section. As discussed in [3, 4] the tapes are paired with their magnetic substrates facing each other and the additional Ni5%W substrates are placed between each pair. The sub-cable design is shown in Fig. 2. Excluding theG-10 holder and cryogenic pipe the active width of the HTS stack is 8 mm.

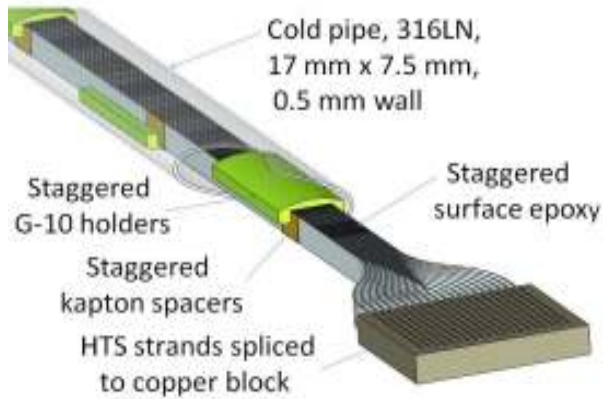

Fig. 2 Arrangement of HTS sub-cable: HTS stack with cryogenic elliptical pipe and HTS tape ends inside the splicing copper block connecting to lead.

The stack is held as an assembly with the help of staggered G-10 holders and epoxy (LORD-305). The epoxy is only touching the top and the bottom surfaces of the stack. The G-10 holders are shaped so their outer surface matches the inner surface of the elliptical pipe. This arrangement prevents HTS stack from twisting. The elliptical pipe is squeezed by the jaws along its entire length to facilitate the HTS stack insertion. Once the HTS stack is properly placed the jaws are released and the spring force of the pipe firmly holds the stack.

The HTS and Ni5%W magnetic substrate tapes are separated from each other with 50 μm thick and 6 mm wide staggered Kapton tape spacers allowing liquid helium coolant to contact directly about 50% surface of the HTS and Ni5%W tapes.

As there are total of 12 elliptical cold pipes in a magnet cable their contribution to dynamic power loss is substantial. For this reason we chose a 316LN SS pipe of 0.5 mm wall. A circular pipe would provide sufficient safety margin for 3 bar pressure of liquid helium coolant, but the elliptical pipe must be reinforced. Using the ANSYS analysis we found that a 1 mm wide spacer placed between two elliptical pipes along their entire length brings the pipe strength back to that of the circular one. The arrangement of the spacer is shown in Fig. 3. The spacer is much wider than 1 mm to facilitate its placement. One side the spacer matches shape of two neighboring pipes and on other side epoxy is used to hold spacer in place. The shape of the pipes was measured after the HTS stacks were installed and then the spacers were fabricated using ABS printer. The applied epoxy was the LORD-305.

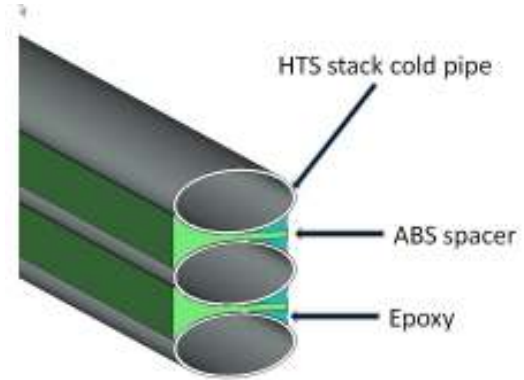

Fig. 3 ABS reinforcing spacers for cryogenic elliptical pipes.

The connection of multiple HTS strands to power lead is the most difficult part of the HTS cable assembly. The joint of HTS strands with the power lead takes place in a cylindrical copper block to which both HTS sub-cables and copper rods of are soldered. On the cable side, as shown in Fig. 4, the splicing block has 6 cavities to house the individual splicing blocks of each sub-cable. These small splicing blocks have 20 slots each to house the HTS strands. Slots are 25 mm long and they are slightly oversized in width to allow the melted solder freely flow to cover both sides of the strand. The strands are guided into the slots with a help of ABS printed supports as shown in Fig. 5

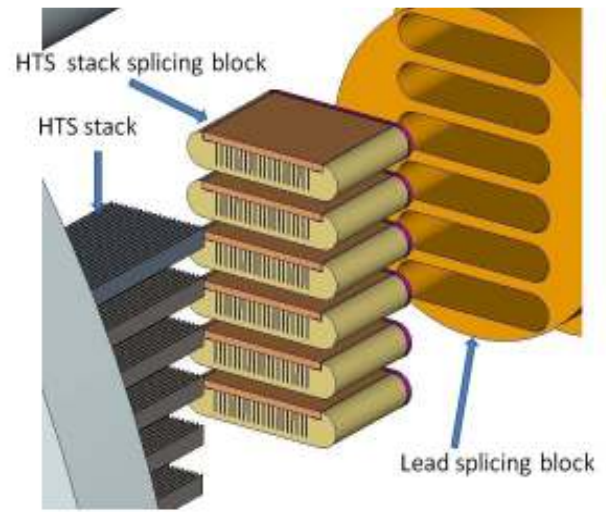

Fig. 4 Arrangement of splicing block for HTS strands.

On other side of the splicing block there are 15 mm deep wells for each of the 32 lead rods. Both HTS strands and lead rods are spliced using the low temperature solder of $120^0$ C. The temperature controlled aluminum heater rings around the connecting splicing block is used to simultaneously solder strands of six sub-cables and 32 rods of the power lead in the splicing block.

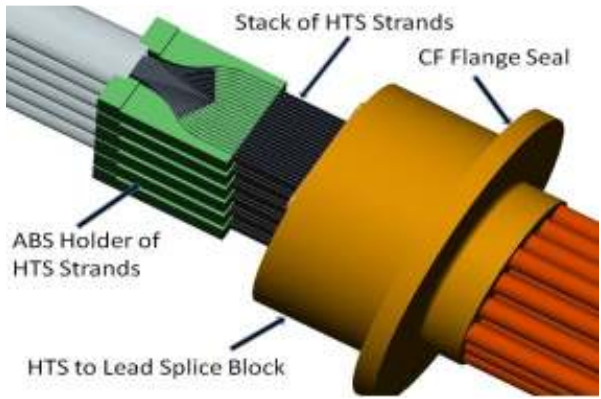

Fig. 5 Connection of HTS sub-cable stacks to power lead.

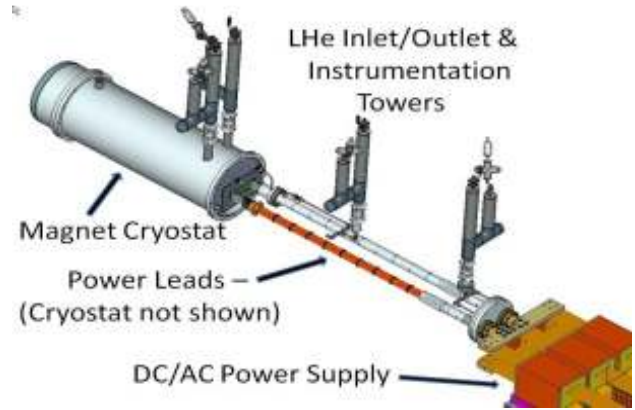

Fig. 8 Cryogenic system; leads cryostat is hidden to show lead structure.

The splicing connection also serves as a separator of cable and lead liquid He flows. The ring in the splicing block is used as a seal for the CF flange. The arrangement of the cryogenic connection to the cable and leads is shown in Fig. 6.

The conventional power leads are scaled-down from the earlier design [5]. Each lead is constructed with 32 copper rods of ¼" diameter and 160 cm length. The rods are held together with the help of G-10 baffles which also force helium to flow up/down through the entire length of the lead to secure even distribution of cooling helium flow.

The assembled magnet coil with the ABS reinforcing spacers is shown in Fig. 9 and the installment of a completed coil into the magnet core is shown in Fig. 10. The magnet core with its power cable installed is shown in Fig. 11.

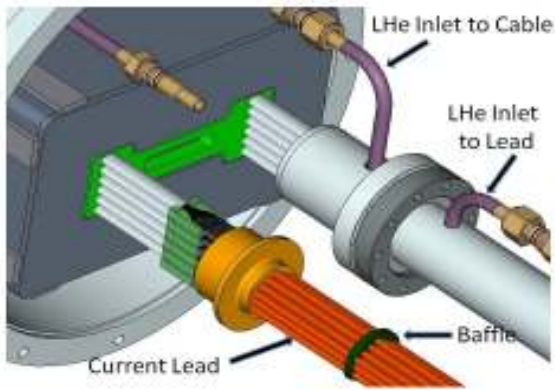

Fig. 6 Arrangement of magnet cable and leads liquid helium flow separation.

### III. Test Arrangement and Assembly Status

A simplified schematic of test arrangement is shown in Fig. 7, and a sketch of the engineering design in Fig. 8. The magnet cable and power leads are cooled with the independent flows of 7 K, 3 bar pressure single-phase liquid helium. The magnet core is warm but its placement in the vacuum pipe provides the cryostat for the superconducting cable as there is no enough space for a cryostat inside the magnet cable cavity.

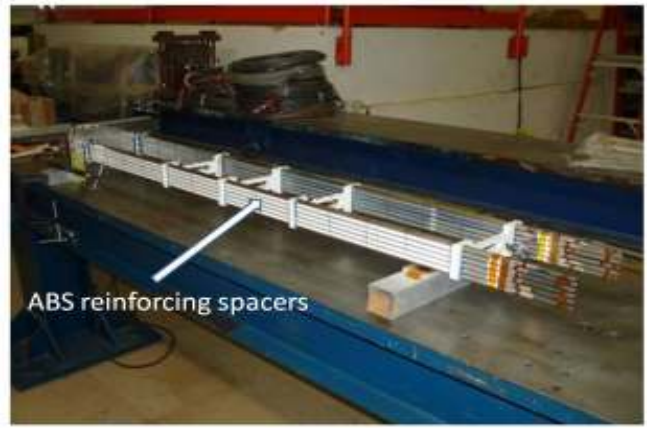

Fig. 9 Magnet power cable with ABS spacers strengthening cold pipes.

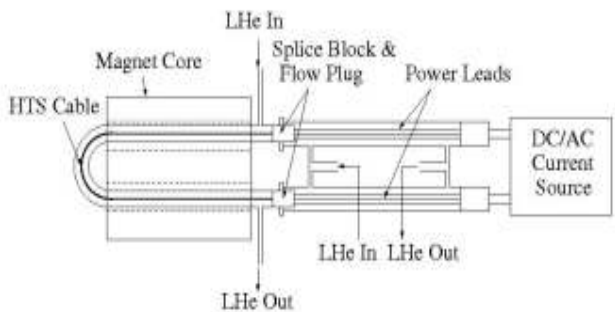

Fig. 7 Simplified schematic of magnet test arrangement.

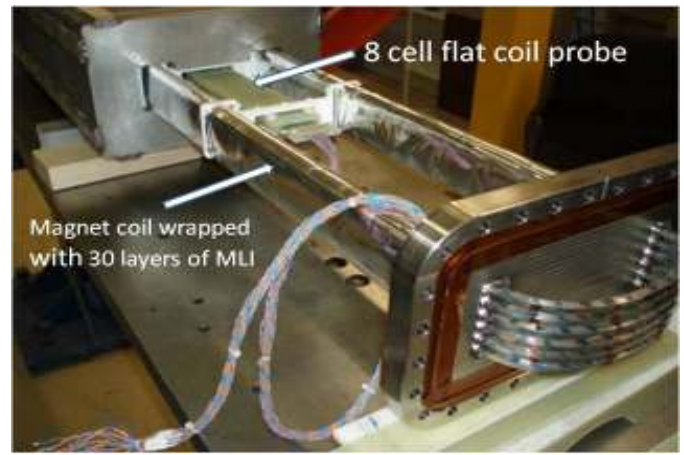

Fig.10 MLI wrapped magnet coil with flat coil probe being inserted into core.

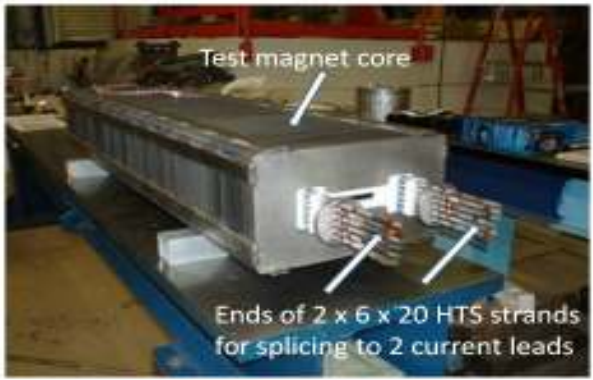

Fig. 11 Coil inside core and HTS strand ends to be spliced to current leads.

All components of magnet test system were fabricated and the assembly work is in progress. The supercritical helium temperature, pressure and flow rate will be measured for both the magnet cable and the current leads. For the test magnet operation the projected temperature margin is 15 K allowing use of temperature measurements along cable path for quench detection. The magnetic field will be measured using the 8-cell flat coil probe of 520 mm length, as shown in Fig. 10.

## IV. POWER LOSS PROJECTIONS AND POSSIBLE APPLICATIONS IN HIGH-ENERGY ACCELERATORS

The power losses of the 344-2G HTS sub-cable in a fast-cycling magnetic field were measured in [4]. We use these data to project power losses for the test magnet and to extend them for the Proton and Muon synchrotrons of the MAP proposal [6]. The accelerator and magnet parameters together with the projected power losses are given in Table I.

TABLE I PROJECTED POWER LOSS FOR PROTON AND MUON SYNCHROTRONS

| Accelerator and Magnet Parameters | | Proton Synchrotron 8 GeV | Muon Synchrotron 63 GeV | 375 GeV |
|---|---|---|---|---|
| Circumference | [m] | 830 | 6860 | |
| $E_{inj}$ / $E_{extr}$ | [GeV/GeV] | 1/8 | 5/63 | 20/375 |
| $B_{inj}$ / $B_{extr}$ | [T/T] | 0.05/0.5 | 0.02/0.24 | 0.08/1.45 |
| Beam gap | [mm] | 40 | 25 | |
| $I_{inj}$ / $I_{extr}$ | [kA/kA] | 2.5/25 | 0.07/7.5 | 0.28/45 |
| $N_{turns}$ | | 1 | 1 | |
| Cycle period | [μs] | 2.77 | 22.9 | 22.9 |
| Cycles to top energy | | 6000 | 12 | 71 |
| Time to top energy | [ms] | 16.6 | 0.275 | 1.63 |
| dB/dt (beam gap & core) | [T/s] | 30 | 840 | |
| dB/dt (cable gap) | [T/s] | 0.3 | 8.4 | |
| Beam rep. rate | [Hz] | 60 | 60 | |
| Superconductor | [HTS] | 344C-2G | 344C-2G | |
| N strands / magnet cable | | 72 | 36 | 180 |
| Cable power loss @ 5K | [W/m] | 1.4 | 30 | 150 |
| Duty factor (DF) | [%] | 1.67 | 1.67 | |
| Cable power loss @ DF | [W/m] | 0.023 | 0.5 | 2.5 |
| Magnet filling factor | [%] | 70 | 70 | |
| Total cable power loss @ 5K | [kW] | 0.6 | 6.5 | 16 |
| Total core power loss @ RT | [kW] | 0.005 | 160 | |

In the HTS cable test [4] the measured power loss was 8 W at 20 T/s cycling rate for a stack of 20 HTS tapes at $8^0$ relative to the cycling 0.5 T field. Our studies indicate that in the optimal position of the cable within the core gap (e.g. position B in Fig. 1) the average field throughout cable space is about 480 G, or 2.8 %, of the 1.7 T field in the magnet beam gap. This value of the field is primarily attributed to the self-field as the progression of the field within the cable space indicates. In the HTS stack arrangement, however, the Ni5%W magnetic substrates are expected to strongly suppress self-field within each stack. The Ni5%W saturates at about 70 G and it may not cover perfectly the HTS strand surfaces. Consequently, the self-field suppression with magnetic substrates may not be 100% effective. Assuming that the self-field is suppressed only by 70% the average field crossing the HTS stack would be then in the range of ~1% of the field in the magnet beam gap. At the 20 T/s rate (and higher) the eddy losses are strongly dominant following the $(dB/dt)^2$ rule. This would lead to a power loss of about 0.8 mW for the cable tested in [4], but with 0.5 T field induced by the cable current itself. Using this assumption we make tentative projections for the power losses for proposed synchrotrons of the Muon Accelerator Program shown in Table 1.

The cable power losses are scaled linearly with number of strands in the cable and with $(dB/dt)^2$ as the eddy losses are strongly dominant for both the 8 GeV Proton Synchrotron and 63 GeV and 375 GeV Muon Synchrotrons. The estimated power losses indicate that they can be acceptable in practical implementation, e.g. the total cable power loss for the 375 GeV Muon Synchrotron may be as low as ~16 kW. This is less than 24 kW required for the Tevatron operating in the same size accelerator ring.

## V. SUMMARY AND CONCLUSIONS

We presented design, construction and test arrangement for a fast-cycling accelerator magnet using the HTS strands to construct the power cable. If tests confirm expectations for magnet power losses, than the proposed HTS-based magnets will constitute rather viable option for the Proton and Muon Synchrotrons of the Muon Accelerator Program at Fermilab.

## ACKNOWLEDGMENT

We are grateful to Frank McConologue, Clark Reid and Julie Kurnat for engineering system design, and Phil Gallo for very meticulous assembly work. We would like also to thank Tom Nicol for performing insightful ANSYS analysis.